\documentclass[a4paper]{jpconf}
\usepackage{graphicx}
\begin{document}
  \newcommand {\nc} {\newcommand}
  \nc {\beq} {\begin{eqnarray}}
  \nc {\eeq} {\nonumber \end{eqnarray}}
  \nc {\eeqn}[1] {\label {#1} \end{eqnarray}}
  \nc {\eol} {\nonumber \\}
  \nc {\eoln}[1] {\label {#1} \\}
  \nc {\ve} [1] {\mbox{\boldmath $#1$}}
  \nc {\ves} [1] {\mbox{\boldmath ${\scriptstyle #1}$}}
  \nc {\mrm} [1] {\mathrm{#1}}
  \nc {\half} {\mbox{$\frac{1}{2}$}}
  \nc {\thal} {\mbox{$\frac{3}{2}$}}
  \nc {\fial} {\mbox{$\frac{5}{2}$}}
  \nc {\la} {\mbox{$\langle$}}
  \nc {\ra} {\mbox{$\rangle$}}
  \nc {\eq} [1] {(\ref{#1})}
  \nc {\Eq} [1] {Eq.~(\ref{#1})}
  \nc {\Ref} [1] {Ref.~\cite{#1}}
  \nc {\Refc} [2] {Refs.~\cite[#1]{#2}}
  \nc {\Sec} [1] {Sec.~\ref{#1}}
  \nc {\chap} [1] {Chapter~\ref{#1}}
  \nc {\anx} [1] {Appendix~\ref{#1}}
  \nc {\tbl} [1] {Table~\ref{#1}}
  \nc {\fig} [1] {Fig.~\ref{#1}}
  \nc {\ex} [1] {$^{#1}$}
  \nc {\Sch} {Schr\"odinger }
  \nc {\flim} [2] {\mathop{\longrightarrow}\limits_{{#1}\rightarrow{#2}}}
  \nc {\textdegr}{$^{\circ}$}
  \nc {\IR} [1]{\textcolor{red}{#1}}
  \nc {\IG} [1]{\textcolor{green}{#1}}

\title{Near-Far Description of Elastic and Breakup Reactions of Halo Nuclei}


\author{\underline{Mahir Hussein}\footnote{Speaker}}

\address{Instituto de F\'isica, Universidade de S\~ao Paulo C.P. 66318, 05315-970 S\~ao Paulo, S.P., Brazil}

\ead{hussein@if.usp.br}

\author{Pierre Capel, Daniel Baye}

\address{Physique Nucl\'eaire et Physique Quantique (CP 229), Universit\'e Libre de Bruxelles, B-1050 Brussels, Belgium}

\ead{pierre.capel@ulb.ac.be, dbaye@ulb.ac.be}

\begin{abstract}
The angular distributions for elastic scattering and breakup of halo nuclei are analysed
using a near-side/far-side decomposition within the framework of the dynamical eikonal approximation.
This analysis is performed for $^{11}$Be impinging on Pb at $69A$MeV.
These distributions exhibit very similar features.
In particular they are both near-side dominated, as expected from Coulomb-dominated reactions.
The general shape of these distributions is sensitive mostly to the projectile-target interactions,
but is also affected by the extension of the halo.
This suggests that the link between elastic scattering and a possible loss of flux towards
the breakup channel is not obvious.
\end{abstract}

\section{Introduction}
The development of radioactive-ion beams in the mid-80s has enabled the exploration of the
nuclear landscape away from stability. This technical breakthrough led to the discovery
of exotic nuclear structures such as hal\oe s \cite{Tan85b,Tan85l}.
Halo nuclei are light neutron-rich nuclei, which exhibit a matter radius significantly larger than their isobars.
This large size is qualitatively understood as resulting from their
small binding energy for one or two neutrons \cite{HJ87}:
Due to their loose binding, these valence neutrons can tunnel far away from
the core of the nucleus and exhibit a large probability of presence at a large distance
from the other nucleons. Halo nuclei have thus a strongly clusterised structure:
they can be seen as a core to which one or two neutrons are loosely bound.
These valence neutrons hence form a sort of diffuse halo around a compact core.
The best known halo nuclei are $^{11}$Be and $^{15}$C, with a one-neutron halo,
and $^6$He and $^{11}$Li, with a two-neutron halo.
Proton hal\oe s can also develop around proton-rich nuclei,
such as $^8$B or $^{17}$F.

Since their discovery, halo nuclei have been at the centre of many studies, both
experimental \cite{Tan96,Jon04} and theoretical \cite{AN03,BHT03}.
Due to their short lifetime, they cannot be studied with usual spectroscopic techniques,
and one must resort to indirect methods to deduce information about their structure.
Reactions are the most used tools to study halo nuclei. In particular, elastic scattering
\cite{Mat06,Dip10} and breakup \cite{Fuk04,BCG05} convey interesting information
about the structure of the projectile.

Recent experimental \cite{Dip10} and theoretical \cite{Mat06} studies of elastic scattering
indicate a strong coupling between scattering and breakup.
On the experimental side, the elastic scattering cross section for $^{11}$Be on Zn around
the Coulomb barrier is significantly reduced at large angles compared to that of non-halo
Be isotopes \cite{Dip10}. One explanation of this unexpected reduction is the transfer of
probability flux from the elastic channel to the breakup channel.
On the theoretical side, Matsumoto \etal have shown that for the elastic scattering of
$^6$He on Bi at low energy CDCC calculations agree with experimental data only if the
breakup channel is included in the model space.

To better investigate the interplay between elastic scattering and breakup, we analyse
theoretically the angular distributions for the elastic scattering and breakup of halo nuclei
within a near/far decomposition \cite{CFH85,HM84}.
We choose $^{11}$Be, the archetypical one-neutron halo nucleus, as testcase and
consider its collision on Pb at $69A$MeV, which corresponds to the conditions of the
RIKEN experiment \cite{Fuk04}.
The calculations are performed within the Dynamical Eikonal Approximation (DEA) \cite{BCG05,GBC06},
which is in excellent agreement with various experimental results.

After a brief reminder of the DEA and the near/far decomposition,
we apply this technique to the elastic-scattering cross section (\Sec{el}).
We then move to the analysis of the angular distribution for breakup
(\Sec{bu}) and show how similar both cross sections are at intermediate energies.
In \Sec{conclusion}, we emphasise the consequences of this analysis for
the study of halo nuclei and provide the prospects of this work.

\section{Theoretical framework}\label{theory}
\subsection{Dynamical eikonal approximation}\label{dea}
Most of the models of reactions involving one-neutron halo nuclei
rely on a three-body description of the colliding nuclei \cite{AN03,BC12}:
a two-body projectile $P$ made up of a fragment $f$ loosely bound to a core $c$
impinging on a structureless target $T$.
The two-body structure of the projectile is described by the phenomenological
Hamiltonian
\beq
H_0=-\frac{\hbar^2}{2\mu_{cf}}\Delta_{\ve{r}}+V_{cf}(\ve{r}),
\eeqn{e1}
where $\ve{r}$ is the $c$-$f$ relative coordinate, $\mu_{cf}$ is the $c$-$f$ reduced mass,
and $V_{cf}$ is a real potential adjusted to reproduce the binding energy of the fragment
to the core and some of the excited states of the projectile.
This potential usually exhibits a Woods-Saxon form factor and may include a spin-orbit coupling term.

The interaction between the projectile components $c$ and $f$ and the target $T$ are
simulated by the optical potentials $V_{cT}$ and $V_{fT}$, respectively.
Within this three-body framework, studying reactions involving one-neutron halo nuclei
reduces to solve the three-body \Sch equation
\beq
\left[-\frac{\hbar^2}{2\mu}\Delta_{\ve{R}}+H_0+V_{cT}(\ve{r},\ve{R})+V_{fT}(\ve{r},\ve{R})\right]
\Psi(\ve{r},\ve{R})=E_{\rm tot}\Psi(\ve{r},\ve{R}),
\eeqn{e2}
where $\ve{R}$ is the $P$-$T$ relative coordinate, $\mu$ is the $P$-$T$ reduced mass and
\beq
E_{\rm tot}=E_0+\frac{\hbar^2K^2}{2\mu}
\eeqn{e3}
is the total energy of the system, with $E_0$ the (negative) energy of the projectile ground state
$\phi_{l_0 j_0 m_0}$ and  $\hbar K$ the initial $P$-$T$ relative momentum.
The quantum numbers $l_0$, $j_0$ and $m_0$ correspond to the $c$-$f$ orbital angular momentum,
the projectile total angular momentum and its projection, respectively.

To describe a reaction in which the halo nucleus $P$ impinges on the target $T$,
\Eq{e2} is solved with the initial condition
\beq
\Psi^{(m_0)}(\ve{r},\ve{b},Z)\flim{Z}{-\infty}e^{iKZ}\phi_{l_0 j_0 m_0},
\eeqn{e4}
where the dependence upon the transverse $\ve{b}$ and longitudinal $Z$ components
of $\ve{R}$ is made explicit.
Equation \eq{e2} must be solved for each value of $\ve{b}$ and of $m_0$.

At sufficiently high incident energy, the eikonal approximation can be performed to ease the
resolution of \Eq{e2}. That approximation consists in assuming that most of the rapid variation
of $\Psi$ in the $P$-$T$ relative coordinate $\ve{R}$ is included in the plane wave $e^{iKZ}$,
i.e. that the three-body wave function is well approximated by that plane wave times a function
$\widehat \Psi$ that does not vary much with $\ve{R}$:
\beq
\Psi(\ve{r},\ve{b},Z)=e^{iKZ}\widehat\Psi(\ve{r},\ve{b},Z).
\eeqn{e5}
Including the eikonal ansatz \eq{e5} within \Eq{e2} leads to
\beq
\left[-\frac{\hbar^2}{2\mu}\Delta_{\ve{R}}-i\frac{\hbar^2 K}{\mu} \frac{\partial}{\partial Z}
+\frac{\hbar^2K^2}{2\mu}
+H_0+V_{cT}(\ve{r},\ve{R})+V_{fT}(\ve{r},\ve{R})\right]
\widehat\Psi(\ve{r},\ve{R})=E_{\rm tot}\widehat\Psi(\ve{r},\ve{R}).
\eeqn{e6}
Since $\widehat\Psi$ varies smoothly with $\ve{R}$, its second-order derivative
$\Delta_{\ve{R}}\widehat\Psi$ can be neglected
in front of its first-oder derivative $K\partial/\partial Z \widehat\Psi$.
Then, considering the energy conservation \eq{e3}, the three-body \Sch equation \eq{e6}
reduces to the DEA equation \cite{BCG05,GBC06}
\beq
i\hbar v \frac{\partial}{\partial Z}\widehat\Psi(\ve{r},\ve{R})
=\left[H_0-E_0+V_{cT}(\ve{r},\ve{R})+V_{fT}(\ve{r},\ve{R})\right]
\widehat\Psi(\ve{r},\ve{R}),
\eeqn{e7}
with $v=\hbar K/\mu$ the initial $P$-$T$ relative velocity.
This equation is mathematically equivalent to a time-dependent \Sch equation
with straight-line trajectories posing $Z=vt$. It can thus been solved using appropriate algorithms,
such as the one described in \Ref{CBM03c}.
However, since the DEA does not assume any semiclassical treatment of the $P$-$T$ relative motion,
$\ve{b}$ and $Z$ are quantal variables. This implies that the DEA includes quantal interferences
such as between different trajectories, which
are missing in time-dependent models \cite{GBC06}.
The DEA therefore significantly improves these models.

The DEA differs also from what is usually called the eikonal approximation \cite{BC12}.
In its usual form, the eikonal approximation includes a subsequent adiabatic approximation to \Eq{e7}
in which the excitation energy of the projectile is neglected, i.e. $H_0-E_0\approx 0$.
In that case, the solution of \Eq{e7} is approximated by the eikonal form factor
\beq
\widehat\Psi^{(m_0)}_{\rm eik}(\ve{r},\ve{b},Z\rightarrow\infty)=e^{i\chi(\ve{r},\ve{b})}\phi_{l_0 j_0 m_0}(\ve{r}),
\eeqn{e8}
where the eikonal phase reads
\beq
\chi(\ve{r},\ve{b})=-\frac{1}{\hbar v}\int_{-\infty}^{\infty}[V_{cT}(\ve{r},\ve{b},Z)+V_{fT}(\ve{r},\ve{b},Z)]dZ.
\eeqn{e9}
The DEA thus improves the usual eikonal approximation by including dynamical effects that are
otherwise neglected. These effects may be very significant such as in Coulomb breakup,
for which the usual eikonal approximation diverges  \cite{GBC06}.

The DEA has been used successfully to describe elastic scattering and breakup of one-neutron
halo nuclei on both light and heavy targets \cite{GBC06}. This approximation has also provided
a reliable description of the Coulomb breakup of the one-proton halo nucleus $^8$B \cite{GCB07}.
More recently, a comparison of various reaction models has shown that the DEA is in excellent
agreement with CDCC at intermediate energies \cite{CEN12} while being much less demanding
on a computational point of view.
The DEA is thus the most efficient model to study reactions involving one-neutron halo nuclei at intermediate energies.
Moreover as it describes simultaneously both elastic scattering and breakup, the DEA is ideal for the present study.

\subsection{Angular distributions and their near/far decompositions}\label{nf}
Within the DEA, the angular distribution for elastic scattering,
i.e. the elastic-scattering cross section, reads \cite{GBC06}
\beq
\frac{d\sigma_{\rm el}}{d\Omega}=K^2\frac{1}{2 j_0+1}\sum_{m_0 m'_0}
\left|\int_0^\infty b db J_{|m'_0-m_0|}(qb)S^{(m_0)}_{m'_0}(b)\right|^2,
\eeqn{e10}
where $J_\mu$ is a Bessel function \cite{AS70}, $q=2K\sin\theta/2$
is the transferred momentum and
\beq
S^{(m_0)}_{m'_0}(\ve{b})=\langle\phi_{l_0 j_0 m'_0} |
\widehat\Psi^{(m_0)}(\ve{b},Z\rightarrow\infty)\rangle-\delta_{m'_0 m_0}.
\eeqn{e11}

To have a better insight into the reaction mechanism that takes place during the scattering of
the projectile by the target, we perform a near/far decomposition of the elastic-scattering
cross section \eq{e10} \cite{CFH85,HM84}.
The idea behind this decomposition is to express the Bessel function as the sum of two
Hankel functions \cite{AS70}:
\beq
J_\mu(z)=\frac{1}{2}\left[H^{(1)}_\mu(z)+H^{(2)}_\mu(z)\right].
\eeqn{e12}
The elastic-scattering cross section can then be expressed as the sum of two terms
obtained by substituting $J_\mu$ by either $H_\mu^{(1)}/2$ or $H_\mu^{(2)}/2$ in \Eq{e10}.
The former is called the \emph{Far side} (F) of the angular distribution \eq{e10},
while the latter is its \emph{Near side} (N):
\beq
\frac{d\sigma_{\rm el}^{\rm F,N}}{d\Omega}=K^2\frac{1}{2 j_0+1}\sum_{m_0 m'_0}
\left|\int_0^\infty b db H_{|m'_0-m_0|}^{(1,2)}(qb)S^{(m_0)}_{m'_0}(b)\right|^2.
\eeqn{e13}
Since these two terms add coherently to form the elastic-scattering cross section,
they may interfere when they reach similar magnitude, which
explains some of the oscillatory patterns observed in angular distributions \cite{HM84}.

\begin{figure}
\center
\includegraphics[width=6cm]{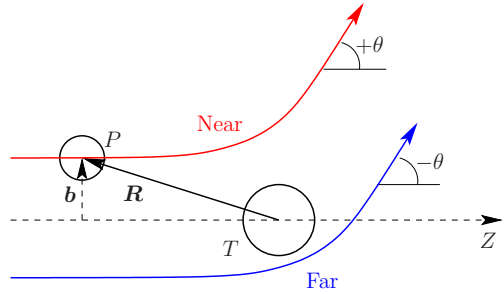}
\caption{Schematic illustration of the Near and Far sides of the angular distribution.}\label{f1}
\end{figure}

The physics behind this decomposition can be understood from the asymptotic
behaviour of the Hankel functions:
\beq
H_\mu^{(1,2)}\flim{z}{\infty}\sqrt{\frac{2}{\pi z}} e^{\pm i(z-\mu\pi/2-\pi/4)}.
\eeqn{e14}
Since $q\approx K\theta$, the N side corresponds to the positive deflection
i.e. repulsive forces (see \fig{f1}). On the contrary the F side carries information
about negative deflection, i.e. attractive forces.

The angular distribution for the breakup of the projectile can also be computed within the DEA
\cite{GBC06}. It corresponds to the breakup cross section expressed as a function of the
scattering angle $\Omega\equiv(\theta,\varphi)$ of the $c$-$f$ centre of mass after
dissociation at a $c$-$f$ relative energy $E=\hbar^2 k^2/2\mu_{cf}$. It reads
\beq
\frac{d\sigma_{\rm bu}}{dEd\Omega}=\frac{2\mu_{cf}KK'}{\pi \hbar^2k}\frac{1}{2 j_0+1}
\sum_{m_0}\sum_{ljm}\left|\int_0^\infty b db J_{|m-m_0|}(qb)S_{kljm}^{(m_0)}(b)\right|^2,
\eeqn{e15}
where
\beq
S_{kljm}^{(m_0)}(\ve{b})=\langle\phi_{k l j m} | \widehat\Psi^{(m_0)}(\ve{b},Z\rightarrow\infty)\rangle,
\eeqn{e16}
with $\phi_{k l j m}$ the continuum wave function describing the broken up projectile in partial
wave $ljm$.

To study the breakup process, we extend the N/F analysis to the angular distribution \eq{e15}.
As for the elastic scattering, the Bessel function is decomposed into the sum of two Hankel
functions \eq{e12}, which provide both N and F sides of the breakup cross section with the
same interpretation, i.e. the contribution to breakup of the repulsive and attractive forces,
respectively.

\section{Elastic scattering}\label{el}
\subsection{{\rm $^{11}$Be} on {\rm Pb} at $69A$MeV}
As a first step in our analysis, we study the elastic scattering of $^{11}$Be on Pb
at $69A$MeV.
As mentioned earlier, $^{11}$Be is the archetypical one-neutron halo nucleus. In our
analysis, it is described as a $^{10}$Be core in its $0^+$ ground state
to which one-neutron is loosely bound.
We choose for the $^{10}$Be-n interaction the potential developed in \Ref{CGB04},
which reproduces the $1/2^+$ ground state in the $1s1/2$ partial wave at 504~keV
below the one-neutron separation threshold. This potential also reproduces the $1/2^-$
bound excited state in the $0p1/2$ orbital and the $5/2^+$ resonance in the $d5/2$
partial wave.
We use the numerical parameters and the potentials $V_{cT}$ and $V_{fT}$ detailed in
\Ref{GBC06}. The numerical technique used to solve the DEA equation \eq{e7} is
explained in \Ref{CBM03c}

\begin{figure}
\center
\includegraphics[width=8cm]{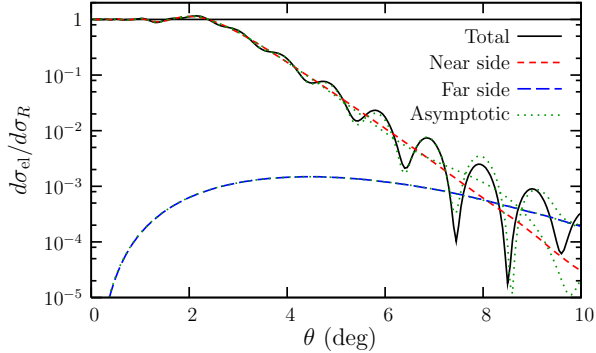}
\caption{N/F analysis of the elastic scattering of $^{11}$Be on Pb at $69A$MeV
\cite{CHB10}.}\label{f2}
\end{figure}

The DEA elastic-scattering cross section is plotted in \fig{f2} as a ratio to Rutherford \cite{CHB10}.
It presents a usual shape with a Coulomb rainbow at about $2^\circ$ followed by
an exponential drop. At larger angles, the angular distribution presents significant
oscillations.
The N/F decomposition shows that at forward angles the process is fully dominated
by the N side, as expected for a (repulsive) Coulomb-dominated reaction \cite{HM84}.
Note that the forward-angles oscillations (i.e. below $2^\circ$) are observed in both
the total cross section and its N side. The N/F interferences therefore cannot explain
this feature of the angular distribution.
At larger angles, i.e. around $8^\circ$, the N and F sides cross, explaining the
oscillatory pattern of the total cross section. This shows in particular that attractive
nuclear forces affect the elastic scattering mostly at large angles, as is expected from
semiclassical models.

The whole interpretation of the N/F decomposition is based on the asymptotic behaviour of
the Hankel functions \eq{e14}. To validate this interpretation, we repeat the calculation of
the cross section, its N and F sides using the asymptotics of the Bessel and Hankel functions.
The angular distributions obtained in this way (dotted lines) are in excellent agreement
with the exact ones, confirming our analysis of this N/F decomposition.

\subsection{Influence of $P$-$T$ interaction}\label{vpt}
\begin{figure}
\center
\includegraphics[width=8cm]{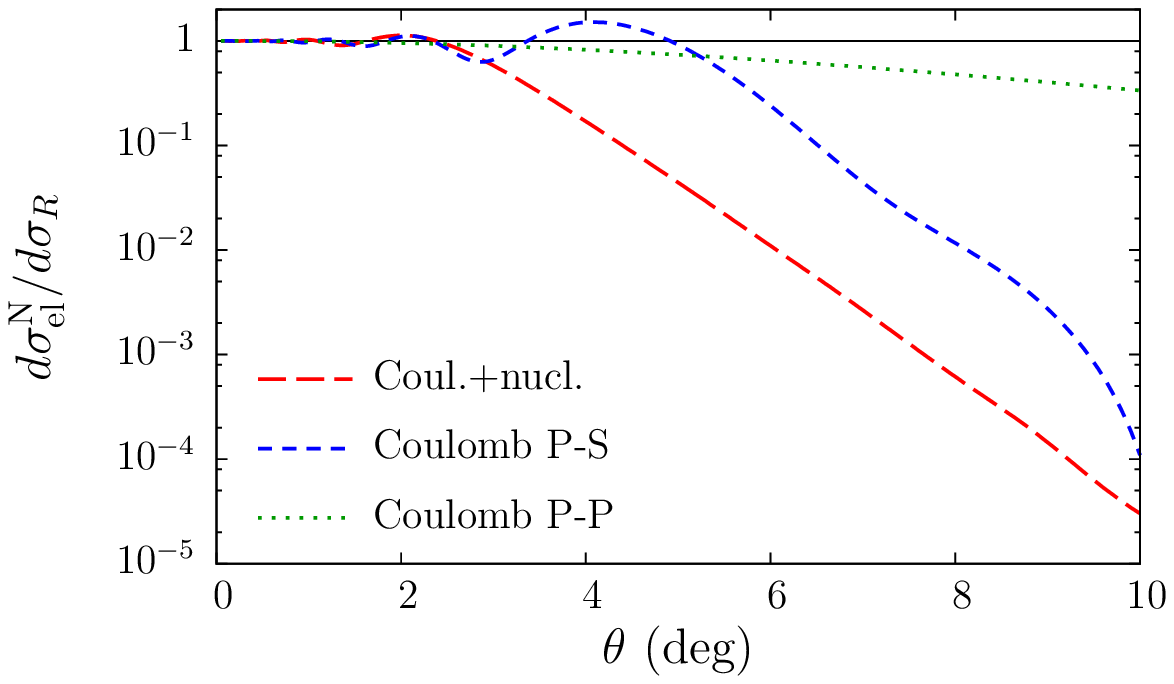}\hspace{-2mm}
\includegraphics[width=8cm]{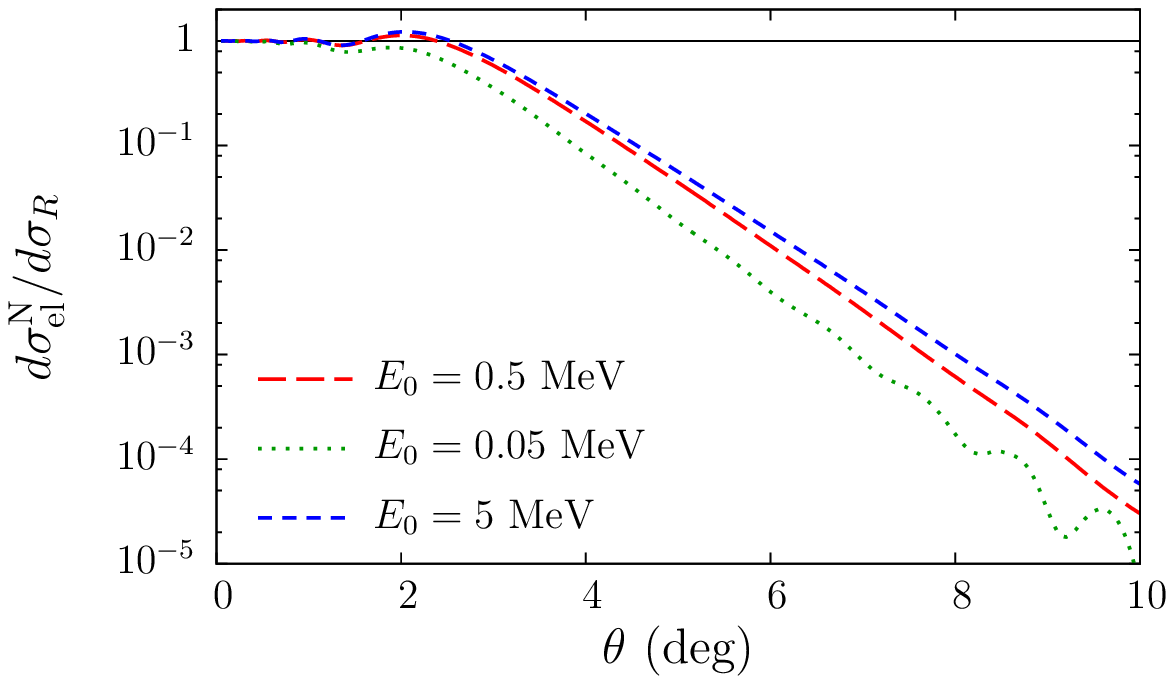}
\caption{Sensitivity of the elastic-scattering cross section to the $P$-$T$ interaction (left)
and the extension of the halo (right).}\label{f3}
\end{figure}

To better apprehend the influence of the choice of $P$-$T$ interaction on the elastic scattering,
we repeat the calculation with the sole Coulomb term of the nuclear optical potential
(i.e. a point-sphere potential, P-S) and a purely point-point Coulomb interaction (P-P).
The dominant N sides of the corresponding elastic-scattering cross sections are plotted in \fig{f3}~(left).
This change of potentials causes dramatic changes in the angular distribution.
It mostly modifies the Coulomb rainbow.
As the full optical potentials, the P-S interaction leads to a Coulomb rainbow,
but its location is shifted from $2^\circ$ to about $4^\circ$.
On the contrary, no Coulomb rainbow is observed with the P-P interaction.
This confirms that the features of the elastic-scattering cross section strongly
depends on the choice of the optical potentials $V_{cT}$ and $V_{fT}$.

Interestingly, the change in the elastic scattering cross section cannot be simply
related to a transfer of flux towards the breakup channel, as postulated in \Ref{Dip10}.
Although the elastic scattering increases at large angles from the Coulomb + nuclear
potential to the purely Coulomb interaction (first with P-S and then even more with P-P),
the total breakup cross section increases as well, as shown in \tbl{t1}.
Since the Coulomb rainbow appears only for the $P$-$T$ potentials that account for
the extension of the projectile and the target (i.e. the full optical potential or just its Coulomb
component P-S), we now analyse the influence of the extension of the halo on these distributions.

\begin{table}
\center
\begin{tabular}{l | c | c | c c c}
\hline \hline
Iteraction & P-P & P-S & \multicolumn{3}{c}{C.+N.}\\ \hline
$|E_0|$ (MeV) & 0.5 & 0.5 & 0.5 & 0.05 & 5\\
$\sigma_{\rm bu}$ (b) & 2.58 & 2.10 & 1.70 & 23.57 & 0.07 \\
\hline \hline
\end{tabular}
\caption{Total breakup cross sections corresponding to the calculations shown in Secs.~\ref{vpt}
and \ref{e0} \cite{CHB10}.}\label{t1}
\end{table}

\subsection{Influence of the size of the halo}\label{e0}
To study the sensitivity of our results to the size of the halo, we repeat the DEA calculations
adjusting the $^{10}$Be-n potential to increase (reduce) the neutron separation energy $|E_0|$
of the $^{11}$Be-like projectile in order to shrink (resp. expand) its halo.
The N side of the elastic-scattering cross section is plotted as a ratio to Rutherford in \fig{f3}~(right).

The slope of the exponential drop after the Coulomb rainbow is sensitive to $E_0$: Reducing
the one-neutron separation energy of the projectile, i.e. expanding its halo, slightly reduces the
elastic-scattering cross section. This effect could therefore be used to get information about the
extension of the halo. However, this dependence remains small in comparison to the influence
of the $P$-$T$ potential [see \fig{f3}~(left)]. There is thus little hope that observing the sole
elastic-scattering cross section could provide unbiased information about the extension of the halo.

Since reducing $|E_0|$ increases the breakup cross section (see \tbl{t1}), we could believe that
the transfer of flux to the breakup channel explains the change in the elastic-scattering cross section.
However, since this increase is much more significant than the drop in the elastic-scattering cross
section, our analysis confirms that there is no direct link between both effects,
as suggested in \Sec{vpt}.

\section{Breakup of {\rm $^{11}$Be} on {\rm Pb} at $69A$MeV}\label{bu}
\begin{figure}
\center
\includegraphics[width=8cm]{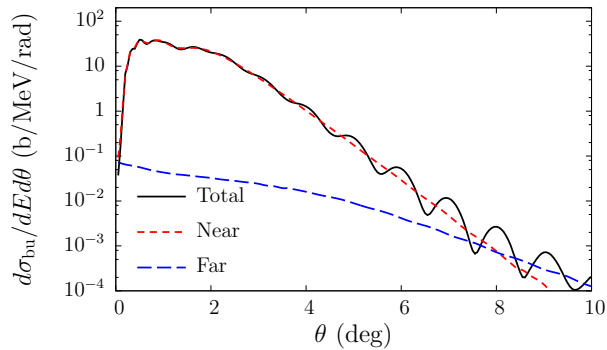}
\caption{N/F decomposition of the breakup angular distribution for $^{11}$Be on Pb at $69A$MeV.
The $^{10}$Be-n relative energy is $E=0.5$~MeV.}\label{f4}
\end{figure}

To better comprehend the link between angular distributions for elastic scattering and breakup,
we perform the same analysis as in \Sec{el} for the breakup cross section \eq{e15}.
For the breakup of $^{11}$Be on Pb at $69A$MeV, the angular distribution (solid line)
and its N (short-dashed line) and F (long-dashed line) sides are plotted as a function of
the scattering angle $\theta$ of the $^{10}$Be-n centre of mass after dissociation \cite{CHB10}
(see \fig{f4}). The $^{10}$Be-n relative energy is $E=0.5$~MeV.

The features of the breakup angular distribution are very similar to those of the elastic-scattering
cross section. First, the full calculation exhibits small oscillations at forward angles before an exponential
drop starting at $2^\circ$, which is reminiscent of the Coulomb rainbow observed in the
elastic-scattering cross section (see \fig{f2}). Second, at forward angles, the breakup is dominated
by its N side, just as in the elastic scattering.
Finally, at larger angles, the total cross section exhibits
oscillations that are explained by  interferences between the N and F side, which cross at about $8^\circ$.

\begin{figure}
\center
\includegraphics[width=8cm]{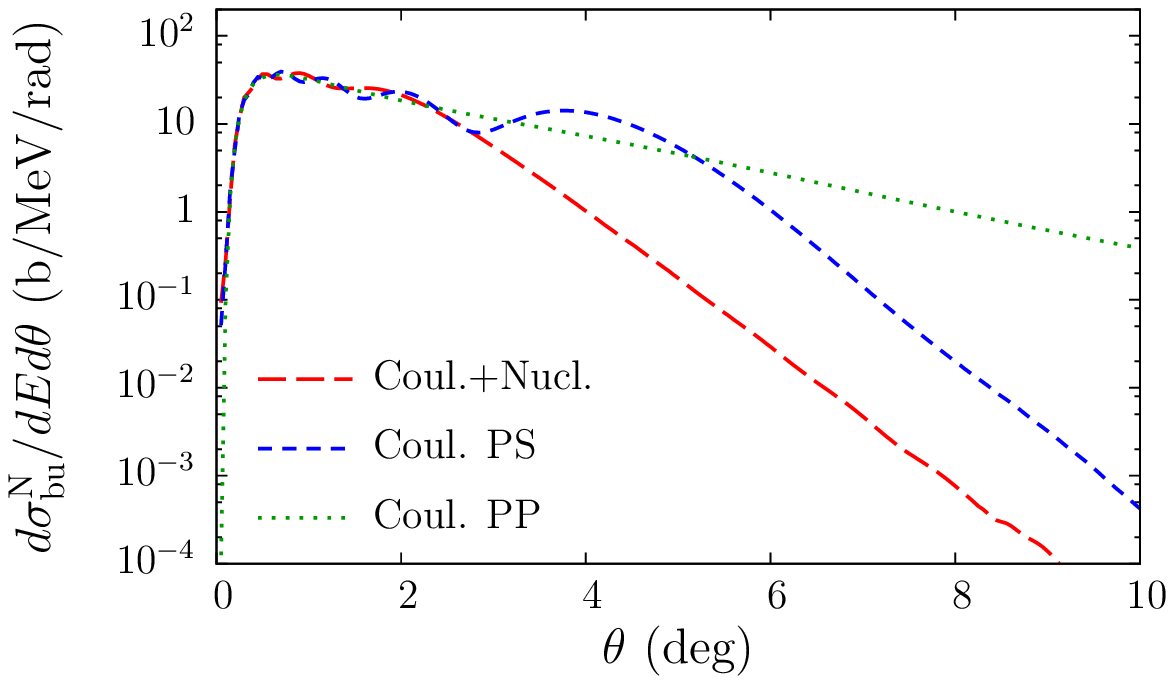}\hspace{-2mm}
\includegraphics[width=8cm]{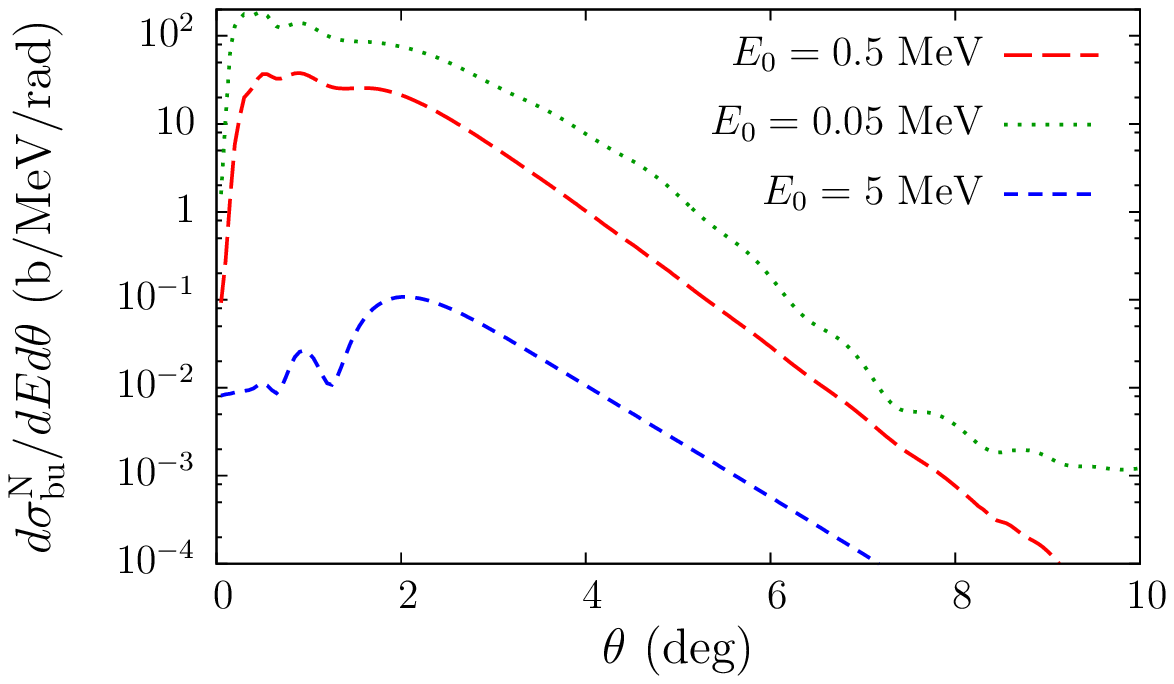}
\caption{Sensitivity of the angular distribution for breakup to the $P$-$T$ interaction (left)
and the extension of the halo (right).}\label{f5}
\end{figure}

\fig{f5} shows the sensitivity of the breakup cross section to the $P$-$T$ potential (left)
and to the binding energy of the neutron $|E_0|$ (right). 
The similarity between elastic scattering and breakup is also observed here.
Using the sole Coulomb part of the optical potentials (P-S) shifts the start of the exponential
drop of the breakup cross section to $4^\circ$, as in the elastic-scattering one [see \fig{f3}~(left)],
and using the purely point-point Coulomb interaction (P-P) leads to no rainbow-like behaviour.
Note that the larger breakup cross section obtained in the P-P case (see \tbl{t1})
is explained by this absence of
Coulomb rainbow in the angular distribution for breakup.
The sensitivity of breakup calculations to the neutron separation energy is also
similar to that observed in the elastic channel. In particular for the slope
of the drop after $2^\circ$, which becomes steeper when $|E_0|$ is reduced [see \fig{f3}~(right)].
This confirms that there is no
direct link between the drop observed in the elastic-scattering cross section and a possible loss of flux
towards the breakup channel since both angular distributions vary in the same way when either
the $P$-$T$ interaction or the projectile structure are modified.

As shown in \Ref{CJN11}, these similarities
can be semi-quantitatively explained within the
Recoil Excitation and Breakup (REB) model developed by Johnson \etal \cite{JAT97}.
Within this adiabatic model, the angular distributions elegantly factorise into the product
of an elastic-scattering cross section for a pointlike projectile and a form factor that accounts
for the extension of the halo \cite{JAT97,CJN11}. The fact that the former appears in both
factorisations and that it contains most of the angular dependence
explains the similarity between the angular distributions.
That analysis also suggests a new observable to study more precisely the halo
structure. As observed in \fig{f3}~(left) and \fig{f5}~(left), the angular distributions are 
very sensitive to the choice of optical potentials.
Since this sensitivity is very similar in both processes,
taking the ratio of two angular distributions removes most of the dependence on the choice
of the $P$-$T$ potentials.
Such a ratio emphasises the nuclear-structure content of the angular distributions \cite{CJN11}.

\section{Conclusion and prospect}\label{conclusion}
In this work, we analyse theoretically elastic-scattering and breakup reactions of halo nuclei
through a N/F decomposition of their angular distributions \cite{CHB10}.
The calculations are performed for $^{11}$Be, the archetypical one-neutron halo nucleus,
impinging on Pb at $69A$MeV, which corresponds to the experimental conditions of \Ref{Fuk04}.
The calculations are performed with the DEA, a reliable and accurate reaction model in which
elastic scatering and breakup are described simultaneously \cite{BCG05,GBC06}.

Our analysis shows that at intermediate energy,
the angular distribution for breakup is very similar to the elastic-scattering cross section:
Both present a Coulomb rainbow at the same scattering angle $\theta$,
they are both N-side dominated at forward angles,
they both exhibit similar sensitivity to the choice of $P$-$T$ interaction
and to the binding energy of the halo neutron.
These results suggest that the projectile is scattered by the target in
a similar way whether it remains bound or it is broken up.
This can be semi-quantitatively understood within the REB model \cite{CJN11,JAT97}.

The present analysis also suggests that
there is no obvious link between the drop observed in the elastic-scattering cross section
and a possible transfer of probability flux towards the breakup channel, as postulated in
\Ref{Dip10}. Since the work of Di Pietro \etal has been performed at lower energy
(around the Coulomb barrier), our conclusions cannot be directly transposed to their study.
A similar analysis within the CDCC framework \cite{DBD10} is planned to see how
elastic scattering and breakup are related to each other at low energy.
Moreover, recent progresses having been made in the modelling of reactions involving
two-neutron halo nuclei \cite{BCD09,PDB12},
an extension of this work for Borromean systems is also planned.

\section*{Acknowledgement}
M.~H. is supported by the Brazilian agencies CNPq and FAPESP.
This text presents research results of the Belgian Research Initiative
on eXotic nuclei (BriX), programme P6/23 on interuniversity
attraction poles of the Belgian Federal Science Policy Office.

\end{document}